\documentclass[manuscript]{aastex}
\usepackage{natbib}
\usepackage{xcolor}
\usepackage{lineno}
\usepackage[figuresright]{rotating}
\usepackage{lscape}
\usepackage{longtable}
\shorttitle{Evolving spectral-timing properties of type-I X-ray bursts from MAXI~J1816--195}
\shortauthors{Chen et al.}

\begin{document}

%%%%%%%%%%%%%%%%%%%%%%%%%%%%%%%Title%%%%%%%%%%%%%%%%%%%%%%%%%%%%%%%%%%%%%

\title{%Simultaneous
%Insight-HXMT and NICER observations on thermonuclear X-ray bursts of MAXI~J1816-195%: burst influence on accretion process
%the hard X-ray deficit $\sim$10\%  detected in a  accretion-powered pulsar
%Broad-band time-resolved spectroscopy on  the outburst and thermonuclear X-ray bursts of MAXI~J1816-195 by Joint observations of  NICER and Insight-HXMT
%The outburst and prolific bursts from the newly discovered millisecond pulsar MAXI~J1816--195 observed by  NICER and Insight-HXMT
%THE PROLIFIC BURSTS BORN OUT OF OUTBURST OF THE NEWLY DISCOVERED MILLISECOND PULSAR MAXI~J1816--195 OBSERVED BY INSIGHT-HXMT ANG NICER
The prolific thermonuclear X-ray bursts from the outburst of the newly discovered millisecond pulsar MAXI~J1816--195 observed by Insight-HXMT and NICER
}

%\author{Y. P. Chen$^{1}$,  S. Zhang$^{1,2}$, S. N. Zhang$^{1,2}$, L. Ji$^{3}$, L. D. Kong$^{1,2}$, P. J. Wang$^{1,2}$}, J. Li,

\author{Yu-Peng Chen\textsuperscript{*}}
\email{chenyp@ihep.ac.cn}
\affil{Key Laboratory for Particle Astrophysics, Institute of High Energy Physics, Chinese Academy of Sciences, 19B Yuquan Road, Beijing 100049, China}

\author{Shu Zhang\textsuperscript{*}}
\email{szhang@ihep.ac.cn}
\affil{Key Laboratory for Particle Astrophysics, Institute of High Energy Physics, Chinese Academy of Sciences, 19B Yuquan Road, Beijing 100049, China}

\author{Long Ji\textsuperscript{*}}
\email{jilong@mail.sysu.edu.cn}
\affil{School of Physics and Astronomy, Sun Yat-Sen University, Zhuhai, 519082, China}

\author{Shuang-Nan Zhang}
%\email{zhangsn@ihep.ac.cn}
\affil{Key Laboratory for Particle Astrophysics, Institute of High Energy Physics, Chinese Academy of Sciences, 19B Yuquan Road, Beijing 100049, China}
\affil{University of Chinese Academy of Sciences, Chinese Academy of Sciences, Beijing 100049, China}

\author{Peng-Ju Wang}
\affil{Key Laboratory for Particle Astrophysics, Institute of High Energy Physics, Chinese Academy of Sciences, 19B Yuquan Road, Beijing 100049, China}
\affil{University of Chinese Academy of Sciences, Chinese Academy of Sciences, Beijing 100049, China}

\author{Ling-Da Kong}
%\email{kongld@ihep.ac.cn}
\affil{Key Laboratory for Particle Astrophysics, Institute of High Energy Physics, Chinese Academy of Sciences, 19B Yuquan Road, Beijing 100049, China}
%\affil{University of Chinese Academy of Sciences, Chinese Academy of Sciences, Beijing 100049, China}
\affil{ Institut f\"{u}r Astronomie und Astrophysik, Kepler Center for Astro and Particle Physics, Eberhard Karls Universit\"{a}t, Sand 1, D-72076 T\"{u}bingen, Germany}

\author{Zhi Chang}
\affil{Key Laboratory for Particle Astrophysics, Institute of High Energy Physics, Chinese Academy of Sciences, 19B Yuquan Road, Beijing 100049, China}

\author{Jing-Qiang Peng}
\affil{Key Laboratory for Particle Astrophysics, Institute of High Energy Physics, Chinese Academy of Sciences, 19B Yuquan Road, Beijing 100049, China}
\affil{University of Chinese Academy of Sciences, Chinese Academy of Sciences, Beijing 100049, China}

\author{Qing-Cang Shui}
\affil{Key Laboratory for Particle Astrophysics, Institute of High Energy Physics, Chinese Academy of Sciences, 19B Yuquan Road, Beijing 100049, China}
\affil{University of Chinese Academy of Sciences, Chinese Academy of Sciences, Beijing 100049, China}

\author{Jian Li}
\affil{CAS Key Laboratory for Research in Galaxies and Cosmology, Department of Astronomy, University of Science and Technology of China, Hefei 230026, China}
\affil{School of Astronomy and Space Science, University of Science and Technology of China, Hefei 230026, China}

\author{Lian Tao}
\affil{Key Laboratory for Particle Astrophysics, Institute of High Energy Physics, Chinese Academy of Sciences, 19B Yuquan Road, Beijing 100049, China}

\author{Ming-Yu Ge}
\affil{Key Laboratory for Particle Astrophysics, Institute of High Energy Physics, Chinese Academy of Sciences, 19B Yuquan Road, Beijing 100049, China}

\author{Jin-Lu Qu}
\affil{Key Laboratory for Particle Astrophysics, Institute of High Energy Physics, Chinese Academy of Sciences, 19B Yuquan Road, Beijing 100049, China}
\affil{University of Chinese Academy of Sciences, Chinese Academy of Sciences, Beijing 100049, China}

%\affil{$^{1}$ Key Laboratory of Particle Astrophysics, Institute of High Energy Physics, Chinese Academy of Sciences, Beijing 100049, China}
%\affil{$^{2}$ University of Chinese Academy of Sciences, Chinese Academy of Sciences, Beijing 100049, China}

%\email{chenyp@ihep.ac.cn, szhang@ihep.ac.cn,jilong@mail.sysu.edu.cn}

%\altaffiltext{5}
%{
%%School of Physical Sciences, Dublin City University, Glasnevin, Dublin 9, Ireland
%}

%%%%%%%%%%%%%%%%%%%%%%%%%%%%%%% Abstract %%%%%%%%%%%%%%%%%%%%%%%%%%%%%%%%%
\begin{abstract}
MAXI~J1816--195 is a newly discovered accreting millisecond pulsar with prolific thermonuclear bursts, detected during its outburst in 2022 June by Insight-HXMT and NICER.
%and returned to quiescent state within a month.
%Jointed observations of the largest soft X-ray band telescope (NICER) and largest hard X-ray band telescope (Insight-HXMT)
%Joint observations of NICER and  Insight-HXMT on its outburst and thermonuclear bursts give us an opportunity to study the interaction between the accretion process and the thermal emission from the Neutron star (NS) surface in a broad band energy of 0.4--100 keV.
During the outburst, Insight-HXMT detected 73 bursts in its peak and decay phase, serving as a prolific burst system found  in the accreting millisecond pulsars.
%The estimated magnetospheric radius from the derived inner-disk radius is comparable with the co-rotation radius.
By analyzing one burst which was simultaneously detected by Insight-HXMT and NICER,   %which is the first reported broad-band simultaneous study of type I bursts using the two telescopes, and find none significant residual from spectral fitting  by a conventional model (blackbody in Xspec),
we find a mild deviation from the conventional blackbody model.
%However, the deviation from the blackbody model turn out to be ubiquitous for other bursters with this burst intensity both for NICER and Insight-HXMT.
%Among the 73 bursts by Insight-HXMT,
By stacking the Insight-HXMT lightcurves of 66 bursts which have similar profiles and intensities, a hard X-ray shortage is detected with a significance of 15.7 $\sigma$ in 30--100 keV. %A cross-correlation analysis between the light curves of the soft and hard X-ray band shows that the corona shortage lags the burst emission by $\sim$ 1 s.
The shortage is about 30\%  of the persistent flux, which is low compared with other bursters. 
%a low value for bursters,
%lower than the other burster systems, 
The shortage fraction is energy-dependent:  larger in a higher energy band.
%Assuming that the accretion column of the magnetic pole is not affected by the burst, this percentage of the deficit is an upper-limit value of the deficit  for the hard X-ray generated from the disk-corona. 
These findings make the newly discovered millisecond MAXI J1816-195 a rather peculiar system compared with other millisecond pulsars and atoll bursters.
%which is a low value compared with the other bursts with this phenomenon.
 %However, this shortage is equivalent to $\sim$ 10\% per-burst/persistent emission in this band, which is much lower than the fraction of the shortage detected in the other bursts with this phenomenon.
%The particular characteristics of the bursts in soft and hard X-ray indicates a  different accretion scenario  from other bursters,  e.g.,  a larger accretion disk radius and two junctions between the accretion flow and the NS surface.
% including but not limited to a larger accretion disk radius or different mechanism of hard X-ray emission.
In addition, based on the brightest burst, %with peak flux $82.6_{-1.5}^{+1.5}$ $10^{-8}~{\rm erg/cm}^{2}/{\rm s}$ detected in the tail of the outburst,
we derive an upper limit of the distance as 6.3 kpc,
%Kong:是否需要给出盘内半径和磁场的估计结果，我觉得得到距离的对于apjl来说可能并不是一个很吸引眼球的结果,
and therefore estimate the upper limit of the inner disc radius of the accretion disc to be $\sim$ 40 km.
Assuming the radius as the magnetospheric radius,
the derived magnetic field strength is about 7.1$\times10^{8}$ G.
%we estimate the magnetic field strength

\end{abstract}

%%%%%%%%%%%%%%%%%%%%%%%%%%%%%%%%% Section 1 %%%%%%%%%%%%%%%%%%%%%%%%%%%%%%%
\section{Introduction}

Accreting Millisecond X-Ray Pulsars (hereafter AMXPs) are
accretion-driven fast-spinning Neutron Stars (NS) with periods shorter than 30 ms (for reviews, see \citealp{DiSalvo2022}).
It is thought that
%the pulse formation process is
the accretion material stripped from the companion is channeled out of the accretion disk through the magnetic line and onto the NS's magnetic poles, which corresponds to the pulse formation process.
%giving rise to X-ray pulsations at the spin frequency.
As a subgroup of $\sim$ 130 low-mass X-ray binaries (LMXBs), since the discovery of the first AMXP  (SAX~J1808.4--3658) in 1998 by RXTE \citep{Wijnands1998, Chakrabarty1998}, roughly two dozen of  AMXPs have been discovered \citep{DiSalvo2022}.
For the X-ray emission during outbursts of AMXPs, the spectral characteristic resembles the hard states of NS LMXBs but with mild spectral evolution, composed of one or two blackbody-like components and an unsaturated Comptonization component with a corona temperature of tens of keV \citep{DiSalvo2022}.

Half of the AMXPs exhibit thermonuclear bursts during their outbursts \citep{Galloway2020}.
Thermonuclear bursts, also named type I X-ray bursts (hereafter bursts), are thermonuclear explosions triggered by unstable burning of accretion material accumulated on the NS surface.
The bursting behavior of AMXPs also resembles other bursts in non-pulsation LMXBs.
It manifests itself as a sudden increase (typically by a factor of 10 or greater) in the X-ray luminosity followed by an exponential decay (for reviews, see \citealp{Lewin, Cumming, Strohmayer, Galloway}).

Since a burst occurs at the NS surface, an interaction \citep{Degenaar2018}  between the burst emission and the neutron star environment should be detected.
In recent ten years, among thousands of observed bursts from the 118  bursters\footnote{https://personal.sron.nl/$\sim$jeanz/bursterlist.html}.
There are  four major impacts observed on the accretion process by bursts, i.e., an enhancement at soft X-ray band, a shortage at hard X-ray band,   a bump peaking at 20--40 keV and/or discrete emission by reflection from accretion disk \citep{int2013, Worpel2013, Ball2004, Keek2014}, and a dip of the persistent emission due to the depleted accretion disk refilling itself \citep{Bult2021}.

For the enhancement at soft X-ray, it is thought that the entire persistent emission level (including disk and corona emission) becomes enhanced by a factor up to ten \citep{Worpel2013, Worpel2015}, alternatively, either the disk or the corona emission brightens \citep{Koljonen2016, Kajava2017}.
These enhancements are observed in  most bright bursts 
%followed by Insight-HXMT, NICER, and AstroSat 
if the bursts are bright enough, 
%i.e., $>$ 1000 cts/s in NICER whole energy band, 
e.g., in 4U~1636--536 \citep{Zhao2022} and Aql~X--1 \citep{Guver2022}.
%(1000--9000 cts/s by NICER).

The shortage during bursts in the hard X-ray of the continuum emission is reported on several bursters, i.e., IGR~J17473--2721 \citep{chen2011, chen2012}, Aql~X--1 \citep{maccarone2003, chen2013}, 4U~1636--536 \citep{ji2013, chen2018, Guver2022a}, GS~1826--238  \citep{ji2014a, Sanchez2020}, KS~1731--260 \citep{ji2014b}, 4U~1705--44 \citep{ji2014b}, 4U~1728--34 \citep{Kajava2017} and 4U~1724--30 \citep{Kashyap2022}, based on RXTE, INTEGRAL, Insight-HXMT, AstroSat and NuSTAR observations.

MAXI~J1816--195, an accretion-powered millisecond pulsar discovered by MAXI in 2022 June \citep{Negoro2022} with peak flux $\sim$ 100 mCrab, spin frequency  528 Hz \citep{Bult2022} and binary orbital period  17402.27 s \citep{Bult2022a}.  Its radio, infrared, and optical counterparts have been identified \citep{Beauchamp2022, Kennea2022, deMartino2022}. %to search of, but not been found yet.%Kong: 此处是否需要补充那几个Atel的结果？
Insight-HXMT and NICER made dense observations on MAXI~J1816--195 until its quiescent state,  where 15 thermonuclear X-ray burst have been detected by NICER \citep{Bult2022b}.

 %\bibitem[Bright et al.  (2022)]{Bright2022}Bright, J., Russell, T., Tremou, E. et al. 2022, ATEL, 15484

In this article, using all the Insight-HXMT observations and   the first 3  observations (public until 2022 June 25) of NICER on MAXI~J1816--195, the broadband energy lightcurves and spectra are studied for both the outburst and bursts.
We derive the burst catalog detected by Insight-HXMT. %since most NICER data are not public.
To assess the burst influence on the persistent emission, we first analyze one burst which was simultaneously detected by Insight-HXMT and NICER; and then, by stacking tens bursts detected by Insight-HXMT, the shortage in the hard X-ray band are given based on the broad-band spectroscopy results.
In the last section, we present our interpretation of these results.
%Kong: 我觉得这段的描述是否应该是具体给出每个Sec几讲了什么？

% of 5 kpc) temperature and a thermal Comptonized spectrum with electron temperature of 30 keV and Thomson optical depth τT ∼ 2for the in the 0.1–200 keV energy band. We have also detected one type I X-ray burst which shows photospheric radius expansion. The burst date from HETE J1900.1–2455, the burst recurrence time is estimated to be about 2 days. No pulsations have been detected either in occurred at an inferred persistent emission level of ∼3–4% of the Eddington luminosity. Using data for all X-ray bursts observed to
%bursts observed with Insight-HXMT and NICER in order to assess the systematic effects in various spectroscopic measurements of their properties. In the first article, we focus on the measurements of the apparent surface areas of 12 neutron stars as inferred from their X-ray spectra during the cooling tails of the bursts. Our goal is to quantify the degree to which (1) the X-ray burst spectra observed in the RXTE energy range can be described by blackbody functions (the so-called color correction arising from atmospheric effects is then applied a posteriori); (2) the entire surface area of each neutron star burns practically uniformly during the cooling of the bursts; and (3) the accretion flows make minor contributions to the emission during the bursts.

\section{Observations and data reduction}

\subsection{Insight-HXMT}
Hard X-ray Modulation Telescope (HXMT, also dubbed as Insight-HXMT, \citealp{Zhang2020}) excels in its broad energy band (1--250 keV) and a large effective area in the hard X-rays energy band.
It carries three collimated telescopes: the High Energy X-ray Telescope (HE; poshwich NaI/CsI, 20–250 keV, $\sim$ 5000 cm$^2$), the Medium Energy X-ray Telescope (ME; Si pin detector, 5–40 keV, 952 cm$^2$) and the Low Energy X-ray telescope (LE; SCD detector, 1–12 keV, 384 cm$^2$).
Under the quick read-out system of Insight-HXMT detectors, there is little pile-up effect   at the burst peak.
Insight-HXMT Data Analysis software (HXMTDAS) v2.05\footnote{http://hxmtweb.ihep.ac.cn/} is used to analyze the data.%Kong: 是否应该给出慧眼软件的网址？

\begin{figure}[t]
\centering
\includegraphics[angle=0, scale=0.8]{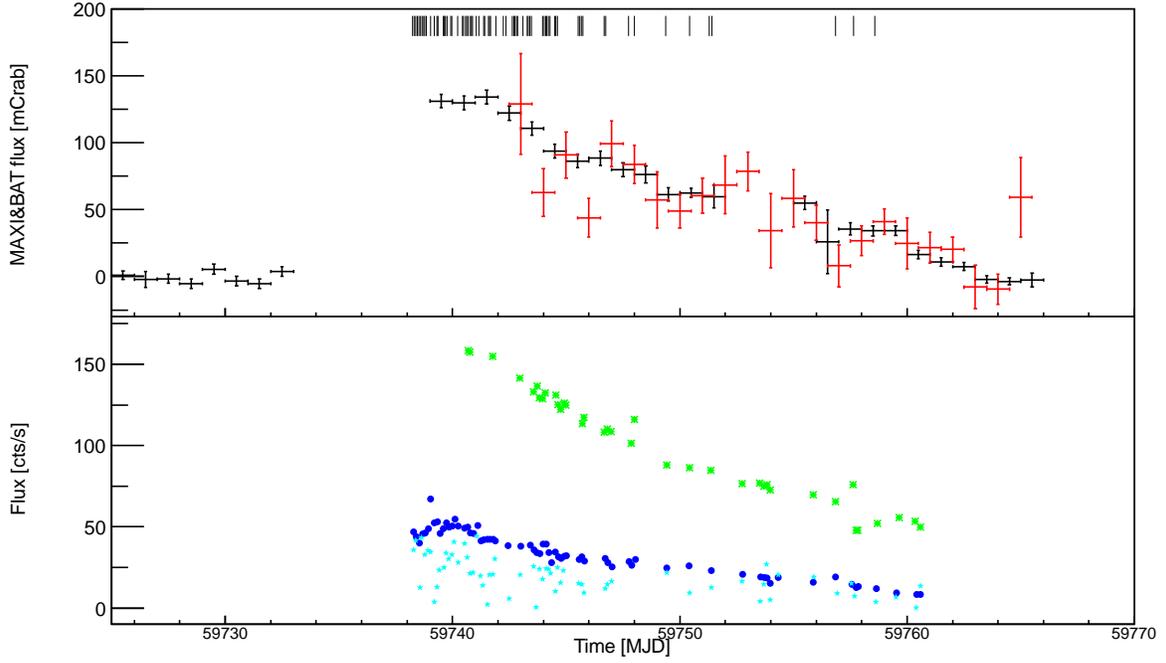}
  \caption{Top panel: daily light curves of MAXI~J1816--195 by MAXI (black) and Swift/BAT (red) during the outbursts in 2022, in 2–20 keV and 15–50 keV, respectively. The 73 bursts   are   indicated by vertical lines. Bottom panel: light curves of MAXI~J1816--195 by LE (green), ME (blue)  and HE (teal), which are rebinned by one obsid ($\sim$ 10000 s).
  }
\label{fig_outburst}
\end{figure}

As shown in Figure \ref{fig_outburst}, starting from 2022 June 8 to June 30, there are 77 observations ranging from P040427500101-20220608-01-01 to P040427502302-20220630-01-01 with a total  observation time of 790 ks on MAXI~J1816--195.
These observations covered the peak and the decay phase of the outburst.
All the observational data above are used in this work.
  Based on the standard pipeline of Insight-HXMT data analysis, the good-time-intervals (GTIs) of LE, ME, and HE are 28 ks, 181 ks, and 146 ks, respectively. 
 
%LE and ME telescopes are both sensitive to burst emission.
We note that the default GTI selection criteria of LE are very conservative because of the influence of light leaks.
To obtain a complete sample of bursts, lightcurves are extracted without filtering GTIs.
Burst-like fluctuations that may be caused by a sharp variation of the background, when the telescope passes the South Atlantic Anomaly (SAA), are excluded.
%when the telescope passes the South Atlantic Anomaly (SAA) and behaves a sharp variation background
%However, LE detectors are easily disturbed by light leaks.
%To avoid missing any burst, we extract the ME lightcurves without screening to extract the burst catalog. {\bf In practice, first, we make a good-time-interval   file  that takes the start time and the stop time of the observation as the start time and the stop time of the good-time-interval file. Then, we use this good-time-interval file to generate the screen file. By the end, the lightcurves and spectra are extracted which are used for the burst analysis. }
%\textcolor{red}{JL: still unclear here.}
%JL: 这部分还是没看,大概能猜出来，但是不敢确定猜的完全对
%If fluctuations of the lightcurve where the bursts appeared are significant, {\bf e.g., when the   telescope passes  the South Atlantic Anomaly (SAA) and behaves a sharp variation background},  the bursts are removed.

%\textcolor{red}{Is it correct?}
As shown in Table \ref{tb_burst_fit}, 73 bursts are found in ME data; among them, 24  are  found in LE data,  and 70   are   found in the HE lightcurves in 20--30 keV with a peak flux of $\sim 70$ cts/s (except for the bright burst, $\sim 400$ cts/s in the whole energy band of HE).
% detected 24 bursts and HE detected 70 bursts. 

\begin{figure}[t]
\centering
\includegraphics[angle=0, scale=0.5]{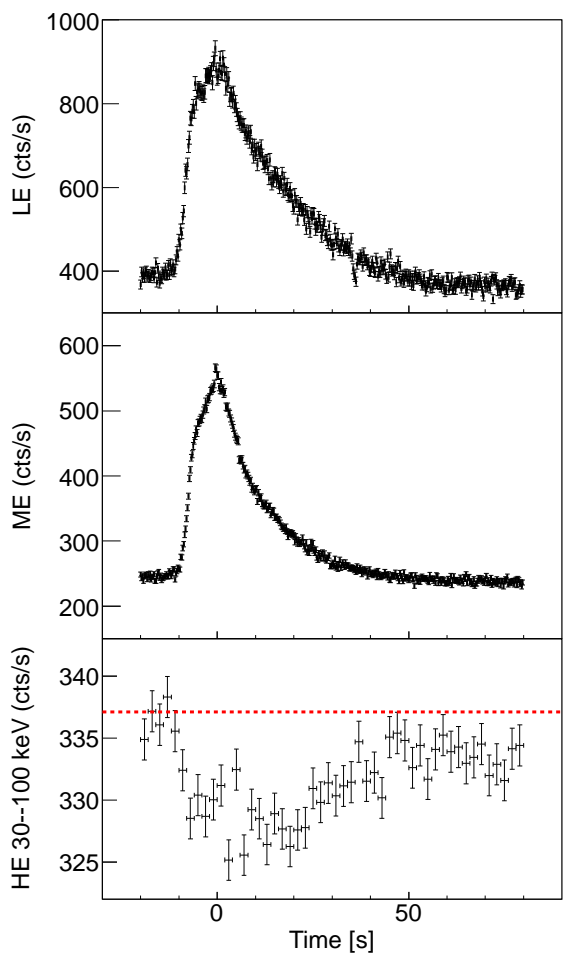}
  \caption{The stacked lightcurves of LE (top), ME (middle), and HE (bottom) of the burst in 1--10 keV, 8--30 keV and 30--100 keV, respectively. The time bin for LE and ME is 0.25 s, for the HE is 2 s. The red line in the bottom panel indicates the pre-burst emission (persistent emission and background) in the HE detectors.  There is one-second data gap around T=35 in LE lightcurve of the burst \#9, which causes a dip around T=35. For other bursts, for LE, ME and HE, there is no such gap  observed.   }
  \label{fig_burst_lc}
\end{figure}

For each burst, we use the time of the ME flux peak as a reference (0 s in Figure \ref{fig_burst_lc}) to produce lightcurves and spectra.
Lightcurves of LE, ME and HE are extracted with a time-bin of 0.25 second. % and  spectra with an time interval  of 1 second are extracted.   from $-$5 s to 45 s
%We divide each burst into intervals of 1 s after the burst onset {\bf (10 second before the burst peak),} and extract the spectra of LE, ME, and HE respectively.
We extract time-resolved spectra of LE, ME and HE with a bin size of 1\,s starting from the onset of each burst  (defined as the time 10 second before the burst peak).
As a conventional procedure, the pre-burst emission (including the persistent emission and the instrumental background) is extracted, which is taken as the background when fitting spectra during bursts.
  In practice, for each burst, we define the time interval between 70 and 20 second before the burst peak as the time window of the pre-burst emission, i.e., [-70 s, -20 s].
%Kong: 是否需要给出关于背景提取的方法描述，LEBKGMAP/MEBKGMAP/HEBKGMAP以及它们的参考文献？

The overlapped observations between Insight-HXMT and NICER are P040427500105-20220608-01-01 and 5533010101, respectively.
Fortunately, these observations were located at the peak of the outburst.
However, based on the recommended procedure of the Insight-HXMT Data Reduction Guide v2.05 \footnote{http://hxmtweb.ihep.ac.cn/SoftDoc.jhtml}, there is no good-time-interval (GTI) of LE.
We loosen the terms of screening criteria for LE data (from private advice from the Insight-HXMT team) with the value of the geomagnetic cutoff rigidity from $>8$ to $>6$; thus, we get 120 s GTI of LE.
The accompanying good-time-interval of ME and HE are 2300 s and 3100 s, which are used for joint spectra fitting of Insight-HXMT and NICER.
% screening criteria.

The other results, e.g, the persistent spectra, background and net lightcurves are obtained  following the recommended procedure of the Insight-HXMT Data Reduction, which are screened with the standard criterion included in Insight-HXMT pipelines: lepipeline, mepipeline and hepipeline. The lightcurves are corrected for the dead time; e.g., the dead time reaches 5\% at the peak of the burst for the HE  lightcurves, which is close to the dead time on the blank sky observations.

For the persistent emission spectral fitting of LE, ME, and HE, the energy bands are chosen to be 2--7 keV and 8--30 keV, and 30--100 keV.
The spectra are rebinned by ftool ftgrouppha \citep{Kaastra2016} optimal binning algorithm with a minimum of 25 counts per grouped bin.

 The LE background model works only in a certain temperature range. This leads to some uncertainties below 2 keV caused by the electronic noise when the temperature exceeds this range after the mid-year of 2019. During a burst with a time-scale of tens of seconds, the temperature fluctuation of LE is so small that can be neglected. 
The resulting electronic noise of the pre-burst spectrum is the same as that of burst spectra.
Therefore, the influence of the electronic noise can be canceled out when we take the pre-burst spectrum as the background of burst spectra.
%Similarly, the difference of the electrical noises between the pre-bursts and the bursts could also be neglected.
In this case, the energy band of LE can be extended to 1--10 keV.
%; due to that taking the pre-burst emission as the background has eliminated the background uncertainty.

The ME energy band used in burst spectral fitting is the same with the persistent emission analysis, i.e., 8--30 keV.
The slices of burst spectra of LE and ME are rebinned by ftool grppha with a minimum of 10 counts per grouped bin, based on the limited photons of the burst slice spectra due to the short exposure time. We added a systematic uncertainty of 1\% to the Insight-HXMT spectra to account for residual systematic uncertainties in the detector calibrations \citep{Li2020}.

\subsection{NICER}

Starting from 2022 June 7, NICER also performed   dense observations on  MAXI~J1816--195.  %We used all of the observational data of MAXI~J1816--195 from NICER publicly  until June 27.
Three obsids (5202820101, 5202820102, 5533010101), from all the NICER public observations of MAXI J1816–195 until June 27, are used in this work.
These observations have a GTI $\sim$ 15 ks and a count rate $\sim$ 600--800 cts/s in the 0.3–12 keV band.   %as shown in Figure  \ref{fig_lc_nicer_le_me}.
Among the three obsids, one burst was detected in 5533010101, with a peak flux of 3018 cts/s.
Fortunately, this burst is also the \#9 burst detected by Insight-HXMT.
%Other characteristics of these bursts will be given in our forthcoming paper.

The NICER data are reduced using the pipeline tool nicerl2\footnote{https://heasarc.gsfc.nasa.gov/docs/nicer/nicer\_analysis.html} in NICERDAS v7a with the standard NICER filtering and using ftool XSELECT to extract lightcurves and spectra.
The background is estimated using the tool nibackgen3C50 \citep{Remillard2022}. The Focal Plane Module (FPM) No. 14 and 34 are removed from the analysis because of increased detector noise.
The response matrix files (RMFs) and ancillary response files (ARFs) are generated with the ftool nicerrmf and nicerarf.
The spectra are rebinned by ftool ftgrouppha \citep{Kaastra2016} optimal binning algorithm  with a minimum of 25 counts per grouped bin.
%Other rebin method, e.g., minimum of 100 counts per grouped bin by ftool grppha, are adapted. As expected, the fit results are consistent with each other within parameter’s error bar.

For the burst detected by NICER, we use the reference time of \#9 burst of Insight-HXMT %to produce the burst spectrum with an interval time of 1 second.
 and divide the burst into intervals of 1 s after the burst onset and extract the spectra.
As a conventional procedure, the pre-burst emission (including the persistent emission and the instrumental background) is extracted as the background when fitting burst spectra,  using the same time interval dealing with the bursts detected by Insight-HXMT: [-70 s, -20 s].
The spectral slices of burst by NICER are rebinned by ftool grppha with a minimum of 20 counts per grouped bin.

The tbabs model with Wilm abundances accounts for the ISM absorption in the spectral model \citep{Wilms2000}.
To erase the residuals in the spectral fitting of the persistent spectra $<$ 1 keV, which is caused by the NICER instrument  and the unmodelled background, the channels $<$ 1 keV are ignored. By the same token, the energy band is extended to 0.4--10 keV during burst spectral fitting.
%Kong: 这两句前后好像有矛盾啊？前面说1 keV需要ignore后面又说用0.4-10 keV。
%three absorption edges are added in spectra fitting: 0.56 keV, 0.71 keV, and 0.87 keV.

 The resulting spectra were analyzed using XSPEC \citep{Arnaud1996} version 12.11.1.
We added a systematic uncertainty of 1\% to the NICER spectrum

%As shown in Figure \ref{fig_lc_nicer_le_me},
%From NICER and Insight-HXMT lightcurves, the none-burst/persistent emission is stable in our observations. We jointly fit the persistent spectra observed with NICER and HMXT, as shown in Figure \ref{sep_nicer}.  During fittings, the LE data in 2--10 keV and ME data in 10--20 keV are used,  while the ME data $>$ 20 keV and the HE spectra are not used to fit  because of faint source flux and strong background emission. The joint fit of the spectra covers an energy band of 0.4--10 keV, 2--10 keV and 10--20 keV for NICER, LE and ME, respectively.

\section{Analysis and Results}
%\subsection{None-burst/persistent spectral fitting on jointed observations of NICER and Insight-HXMT}
\subsection{Fitting the joint Insight-HXMT/NICER spectrum of persistent emission}

%The jointed Insight-HXMT and NICER data in a broader energy range of 1--100 keV allow us to utilize a more physical model to fit the persistent emission, rather than the simplified model, i.e., a simple photon power law, a power law with high energy exponential roll-off (cutoffpl in xspec) and a broken power law (bknpow in xspec).
We fit the joint NICER and Insight-HXMT(LE and ME) spectrum with an absorbed convolution thermal Comptonization model (with an input seed photon spectra diskbb), available as thcomp (a more accurate version of nthcomp) \citep{Zdziarski2020} in XSPEC, which is described by the optical depth $\tau$, electron temperature $kT_{\rm e}$, scattered/covering fraction $f_{\rm sc}$.

% and assumes a spherical distribution of the thermal electrons for the Comptonization.

The  hydrogen column (tbabs in XSPEC) %is fixed at 1.5$\times$10$^{22}~{\rm cm}^{-2}$  \citep{Penninx1989},
accounts for both the line-of-sight column density, as well as any intrinsic absorption near the source.
%and corresponds to the best-fit value for the persistent spectrum.
The seed photons are in the shape of diskbb since the thcomp model is a convolution model, and a fraction of Comptonization photons is also given in the model.
%from any form of the seed spectrum.
%is provided as a convolution function, and thus it can be used with any form of seed photons.  This is a convolution model, allowing for Comptonization of a fraction of photons from any form of the seed spectrum. We assume the seed photons have a disk blackbody spectrum (diskbb; Mitsuda et al. 1984).

%In addition, a diskbb component  .
Normalization constants are included during fittings to take into account the inter-calibrations of the instruments.
We keep the normalization factor of the NICER data with respect to the LE, ME and HE data to unity.
To ease the residuals around 6.4 keV, a Gaussian emission line is added and fixed at 6.4 keV, corresponding to the iron emission line reflected from the disk.
%During the spectral analysis, a 1\% systematic error is added to account for the uncertainties of the background model and  calibration.

%0.4--10 keV
% Using the model above, we find an acceptable fit: reduced $\chi_{\upsilon}$=1.04 (d.o.f. 206; Figure \ref{sep_nicer} and Table  \ref{persist_fit}), with  the inner disc radius $R_{\rm diskbb}$ and  temperature $kT_{\rm in}$ are found to be $\sim 44.6_{-6.8}^{+12.5}$ km (with distance 6.3 kpc and inclination angel 0 degree) and   $0.47_{-0.02}^{+0.03}$ keV respectively.  The thcomp parameters, the electron temperature  $kT_{\rm e}$, optical depth $\tau$ and scattered/covering fraction $f_{\rm sc}$  is  $11.4^{+0.7}_{-0.6}$ keV, $5.19^{+0.23}_{-0.24}$ and  $0.46_{-0.05}^{+0.05}$, respectively.

% 1--10 keV
Using the model above, we find an acceptable fit: reduced $\chi_{\upsilon}$=1.13 (d.o.f. 192; Figure \ref{fig_outburst_spec} and Table \ref{tb_persist_fit}), with the inner disc radius $R_{\rm diskbb}$ and  temperature $kT_{\rm in}$   found to be $\sim 39.6\pm5.1$ km (with a distance 6.3 kpc and an inclination angle 0 degree) and $0.48\pm0.01$ keV respectively. The thcomp parameters of the electron temperature $kT_{\rm e}$, optical depth $\tau$ and scattered/covering fraction $f_{\rm sc}$ are $10.8^{+0.6}_{-0.2}$ keV, $5.41^{+0.25}_{-0.07}$ and $0.48\pm0.01$, respectively.
The derived hydrogen column density $N_{\rm H}$ is $2.37_{-0.03}^{+0.02}\times 10^{22}~{\rm cm}^{-2}$.

% $2.37_{-0.03}^{+0.02}$ & $5.41^{+0.25}_{-0.07}$ & $10.8^{+0.6}_{-0.2}$ & $0.52_{-0.01}^{+0.02}$  & $0.48_{-0.01}^{+0.01}$ & $39.5_{-5.1}^{+5.1}$ & 6.4 (fxd)& $0.86_{-0.14}^{+0.15}$ & $5.4_{-1.1}^{+0.9}$ & 217/192  &$1.11_{-0.01}^{+0.01}$

% {\bf The  derived hydrogen column density  $N_{\rm H}$  is $\sim$1.3$\times 10^{22}~{\rm cm}^{-2}$, which agrees with values previously reported  in a range of 0.9--1.5 $\times 10^{22}~{\rm cm}^{-2}$ \citep{PenninxW1989,ArmasPadilla2017}.
%The thcomp parameters,   $\tau$ and $kT_{\rm e}$ are well consistent with a previous outburst in soft state of 4U 1608--52 \citep{ArmasPadilla2017}, which derived the parameters from the nthcomp model. }
The constants of LE, ME, and HE are 1.33$_{-0.15}^{+0.12}$, 0.97$\pm$0.02 and 0.81$\pm$0.04, respectively.
The inferred bolometric flux in 0.01--1000 keV is  %$1.19\pm0.01\times10^{-8}~{\rm erg/cm}^{2}/{\rm s}$. %0.4--100 keV
$1.11\pm0.01\times10^{-8}~{\rm erg/cm}^{2}/{\rm s}$,
%1--100 keV
corresponding to 28.9\% $L_{\rm Ledd}$ at distance of 6.3 kpc and $L_{\rm Ledd}=1.8\times10^{38}$ erg/s.

The other scenario, i.e., substituting the diskbb component by a blackbody component in the aforementioned convolution model, is also attempted.
Taking this approach, spectral fits yield roughly the same thcomp parameters and reduced $\chi_{\upsilon}$=1.12  (the same d.o.f.).
However, the derived blackbody radius is $86\pm10$ km, which is far greater than the NS radius,  although it is an upper limit considering the  distance used is an upper limit.  
Suppose the derived blackbody radius was comparable with the radius of the NS $\sim 10$ km, the distance adopted  should be $\sim$0.8 kpc, which  is unlikely to occur.
%Even if the distance is 3.15 kpc, the derived blackbody radius is 43 km, which is still greater than the NS radius.  However, the derived luminosity of the outburst is brighter than the Eddington luminosity.}
%\textcolor{red}{JL: but why we assume d=3.15\,kpc? In addition, if the source is close to us, the inferred luminosity should be quite faint. Why "brighter than the Eddington luminosity"?}
%A hybrid model \citep{ArmasPadilla2017}, i.e., three-component model (diskbb+thcomp*bb or bb+thcomp*diskbb) is not attempted, since the above two-component model is able to produce the data.

\subsection{Fitting the joint Insight-HXMT/NICER spectra of burst emission }

%When we fit the burst spectra, we estimate the background using the emission before the burst, i.e., assuming the persistent emission is unchanged during the burst.

\begin{figure}[t]
\centering
\includegraphics[angle=0, scale=0.25]{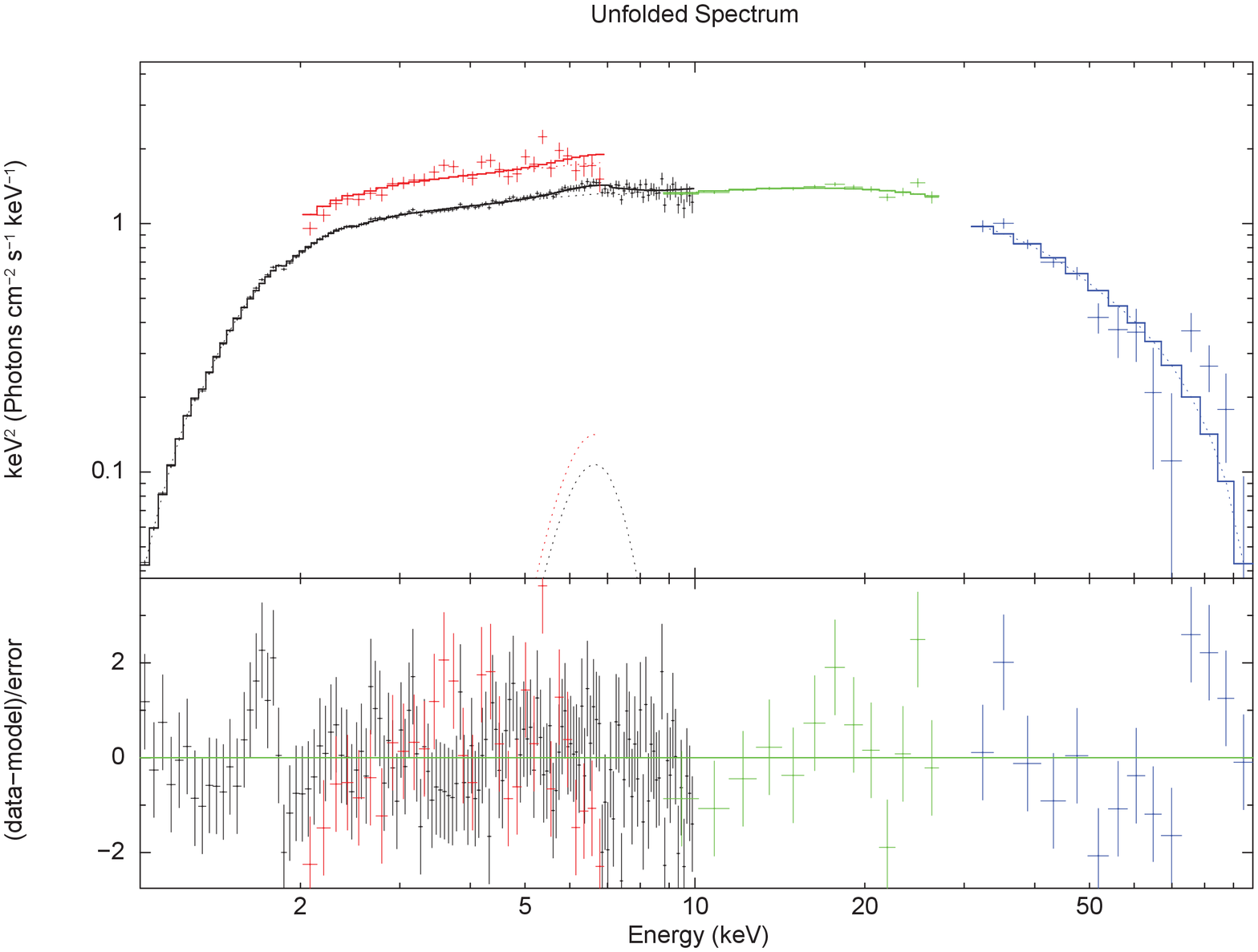}
\includegraphics[angle=0, scale=0.24]{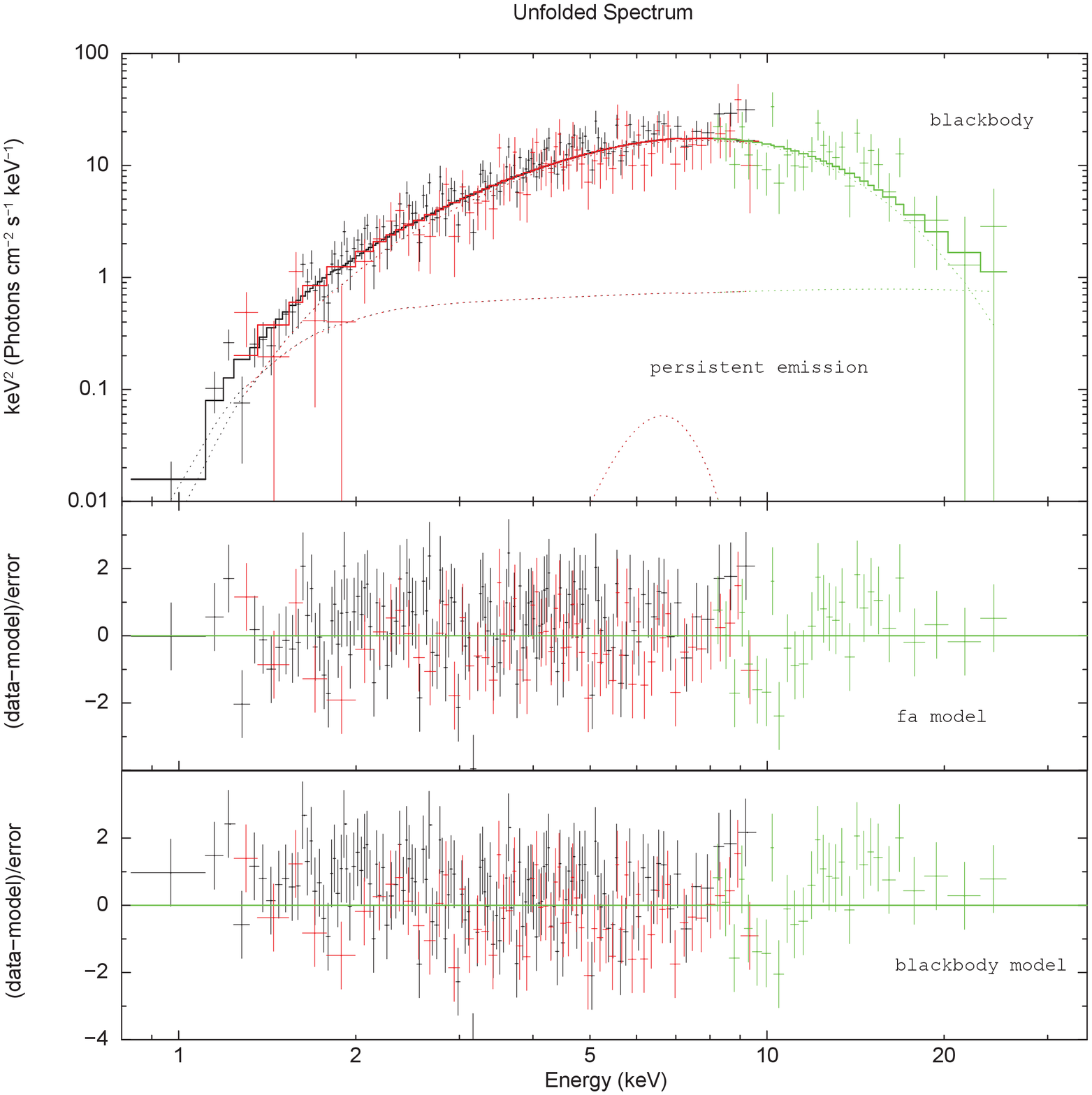}
  \caption{Left panel: Simultaneous broad-band energy spectrum of MAXI~J1816--195  as observed from NICER (black), Insight-HXMT/LE (red), Insight-HXMT/ME (green) and Insight-HXMT/HE (blue); the best-fitting model consists of an absorbed convolution
 thermal Comptonization model (with an input seed photon spectrum diskbb) and an absorbed Gaussian emission line fixed at 6.4 keV.
 %Right panel: the  spectral fitting results by NICER (black), Insight-HXMT/LE (red) and Insight-HXMT/ME when the burst reached its peak flux.
 Right panel: the  spectral fitting results by NICER (black), Insight-HXMT/LE (red) and Insight-HXMT/ME (green) when the burst reached its peak flux with $f_{a}$ model model (top), the blackbody model and the enhancement of the persistent emission are labeled; the two panels below: residuals of spectral fitting results by  $f_{a}$ model (middle) and  an absorbed black-body (bottom).
 }
\label{fig_outburst_spec}
\end{figure}

\begin{figure}[t]
\centering
\includegraphics[angle=0, scale=0.5]{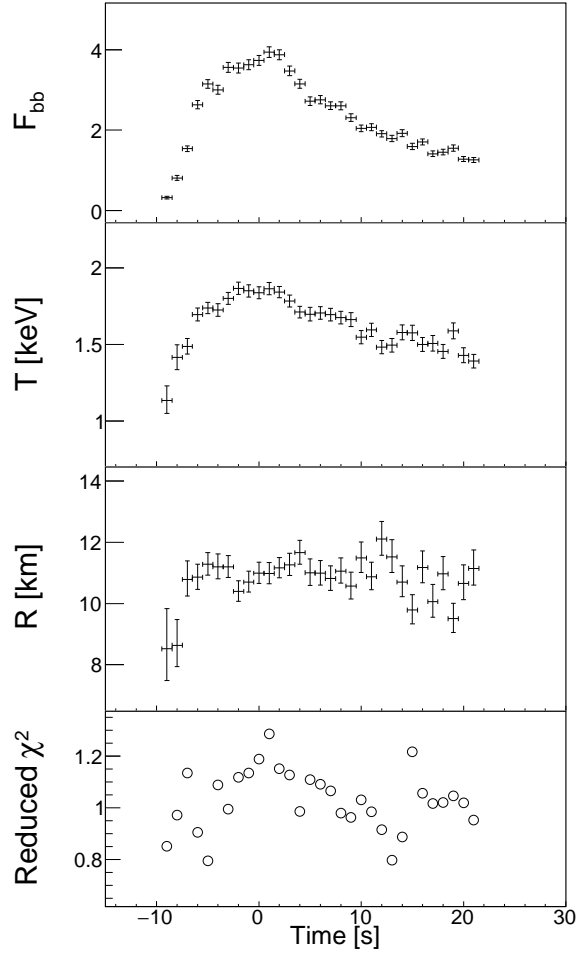}
  \caption{
Spectral fitting result of the burst  \#9 with time bin 1 second  with  an absorbed blackbody (black),
including the time evolution of the blackbody bolometric flux $F_{\rm bb}$, the temperature $kT_{\rm bb}$, the observed radius $R$ of NS surface at 6.3 kpc, the goodness of fit $\chi_{v}^{2}$.
%The black and red indicates the fitting results by a absorbed  blackbody and $f_{a}$ model, respectively.
The bolometric flux of the blackbody model $F_{\rm bb}$ is in units of $10^{-8}~{\rm erg/cm}^{2}/{\rm s}$.   }
\label{fig_burst_fit}
\end{figure}

We follow the classical approach to X-ray burst spectroscopy by subtracting the persistent spectrum and fitting the net spectrum with an absorbed blackbody and a fixed hydrogen column density derived from the fitting result of persistent emission.
As shown in Figure \ref{fig_outburst_spec} and Figure \ref{fig_burst_fit}, such a spectral model generally results in an acceptable goodness-of-fit, with a mean reduced $\chi^{2}_{\upsilon}\sim$0.8--1.2 (d.o.f. 70--200).
%^{2}_{\upsilon}$..  %are above 1.5 (d.o.f. 60--80).
The profiles flux, temperature and radius are characteristic of bursts:
%The flux, temperature, and radius profiles indicate the characteristics of burst:
a flux with fast-rising/exponential-decaying and spectral-softening during decay.
The temperature and radius reach a maximum value of $\sim$2 keV and $\sim$11 km at the burst peak.%Kong 感觉图3中的R更接近13 km?

The $f_{a}$ model is also used to fit the burst spectra.
Following \citet{Worpel2013} we then include an additional component for fitting the variable persistent emission.
We assume that during the burst the spectral shape of the persistent emission is unchanged, and only its normalization (known as a $f_{a}$ factor) is changeable.
%As reported earlier by RXTE and NICER, the $f_{a}$ model provides a roughly same good fit with the conventional one (absorbed blackbody).
We compare the above two models  using the F-test. % As shown in Table \ref{table},
In most cases, the $f_{a}$  model  does not  apparently  improve the fits with a p-value $>3\times10^{-3}$ except the spectrum when the burst reaches its peak flux. %\textcolor{red}{JL: improve or not improve?}
For the spectrum when the burst reaches its peak flux, the $f_{a}$ factor reaches a maximum $0.54\pm0.11$\footnote{ 
%Please not that this is not the traditional definition of the $f_{a}$ model. 
Please note that the definition of $f_{a}$ in our paper is not the same as the traditional one by \citet{Worpel2013}.
Instead, our definition of $f_{a}$=0 indicates that there is no enhancement or deficit of the persistent emission during the burst. 
}, accompanying with the change of the temperature and the radius of the blackbody within a variation of $\sim10\%$.
As shown in Figure \ref{fig_outburst_spec}, at this data point, the $f_{a}$  improves the fits with a p-value 2.3$\times10^{-5}$.

\subsection{Fitting the Insight-HXMT spectra of burst emission }

The same procedure on all bursts is carried out to fit the burst spectra detected by Insight-HXMT.
As shown in Table \ref{tb_burst_fit}, the burst unabsorbed bolometric peak flux $F_{\rm p}$ , burst fluence $E_{\rm b}$ and burst duration $\tau$=$E_{\rm b}/F_{\rm pk}$, are given.

The brightest burst is the last burst \#73, with a maximum peak flux $8.03_{-0.39}^{+0.40}\times10^{-8}~{\rm erg/cm}^{2}/{\rm s}$.
 The burst did not show photospheric radius expansion; assuming the empirical Eddington luminosity of $3.8\times10^{38}$ erg/s \citep{Kuulkers2003}, we derive an upper limit on the source distance of 6.3 kpc.

Based on the  bursts time interval $\sim$ 1.15 hour,
%Based on the total exposure time of all bursts $\sim$ 1.15 hour,
the ratio $\alpha$  of the integrated persistent flux to the burst fluence is $\sim$ 45 (a detailed study on the burst statistical behaviors will be reported elsewhere by Wang et al.).
%which the standard measure of the relative efficiency of the two processes.

\subsection{Stacked lightcurves/spectra of burst by Insight-HXMT}

In the HE lightcurves above 30 keV, we notice there is no significant flux deviation from  the persistent emission during the burst.
  It is caused by  that  the vast majority  of the thermal emission of an X-ray burst lies  outside the HE energy range ($>$ 30 keV).
%Thus is caused by the faint (low-flux) and cold (low-temperature) of individual bursts.
The same as our previous procedures \citep{chen2013}, the burst lightcurves are stacked to improve the statistics.

Table \ref{tb_burst_fit} shows that  most bursts have a peak flux $\sim 4\times10^{-8}~{\rm erg/cm}^{2}/{\rm s}$ and a duration $\sim$ 20 s, except  the last 7 bursts.
Thus, we choose the first 66 bursts to stack since  after that the persistent emission in the hard X-ray band  is $<$ 100 mCrab and
 the latter bursts have a different profile. %, i.e., shorter duration and brighter peak flux.
%latter bursts have a different profile, i.e., shorter duration and brighter peak flux.
Our forthcoming publication will give other features, e.g., burst profiles, intervals, and relation with persistent emission.
With respect to the burst peak time as a reference time, the lightcurves of the 66 bursts are stacked and averaged in each time bin, respectively.

As shown in Figure \ref{fig_burst_lc}, accompanying the burst rise detected by LE\&ME, there is a flux-dropping in the HE lightcurves.
Following the decay of the bursts, the HE flux restores to the pre-burst level.
Considering the pre-burst emission of $\sim$337 cts/s (including $\sim$40 cts/s persistent emission and $\sim$297 cts/s background emission), the HE decrement reaches a maximum of $\sim$ 12 cts/s at the burst peak, which amounts to $\sim$ 30\% of the whole persistent flux in 30--100 keV.
The   average  deficit   from $-$5 s to 45 s (burst peak time as 0 s) is  
$-7.59\pm0.33$ cts/s.
%, amounting to  a detection significance of 23 $\sigma$.

{\bf 
A cross-correlation analysis is performed between LE and HE lightcurves with a bin size of 0.25\,s \citep[for details, see][]{chen2012}. The minimum of the cross-correlation function appears at $\sim$ 1\,s, which indicates that the hard X-ray deficit lags the burst emission with $\sim$ 1\,s.
}

%The cross-correlation between the LE lightcurve and HE lightcurve is $\sim$ 1 s,  which indicates that the hard X-ray deficit lags the burst emission with $\sim$ 1 s. 
%Since the time bin used in the caculation is 1 s, which indicates the

%To study the hard X-ray deficit  as the energy band, we stack the whole burst spectra.
To estimate the deficit significance, another approach based on the stacked spectra is carried out. First, we extract each burst with an exposure time of 50 s, i.e., the interval of 5 s before the burst peak to 45 s after the burst peak. Then, ftool addspec is used to combine the burst spectrum. All the pre-burst spectra with the same exposure time are also stacked.
%Finally, taking the pre-burst spectrum as background, 
%{\bf the net stacked burst spectrum are given.}
%we fit the stacked burst spectra with two blackbodies (absorption is thawed to vary to erase the residuals $<$ 2 keV). As shown in Figure \ref{fig_burst_spec_le_me_he}, the model could fit the spectrum $<$ 30 keV; the deficits $>$ 30 keV are significantly detected. 

Based on the stacked spectra of burst and pre-burst emission,  we get the count numbers for the burst ($N_{\rm b}=971431$ counts) and the pre-burst ($N_{\rm pre}=993402$ counts)  with a same exposure time in 30--100 keV.  
\textbf{Assuming the count numbers follow a Gaussian distribution, the significance of this deficit in 30–100 keV is estimated at $(N_{\rm pre}-N_{\rm b})/\sqrt{N_{\rm pre}+N_{\rm b}}=15.7\sigma$}.
%(993402-971431)/sqrt(971431)=22.3 $\sigma$. 
%We also note that the exposure times in the spectra are not dead-time corrected and that this correction could lead to a reduction of counts by 5\%. Considering the dead-time correction of the two spectra, the  significance of this deficit is revised up to 16.1 $\sigma$, The above analysis shows the significance of the deficit derived from the spectra and lightcurves is consistent.
%\textcolor{red}{I fully agree with the reviewer's comments. More discussions.}

%还不如画，两个能谱，1：burst期间的能谱；2，persistent能谱。两个画在一起，但是别减去。

%We also notice that the deficit becomes less significant with the energy.
%To further estimate the variation of the deficit with the energy, three spectra are used as inputs for estimation, i.e., the detected spectrum of burst (blue points of the 2nd panel of Figure \ref{fig_burst_fake}), the faked spectrum  if there is no deficit (red points of the 2nd panel of Figure \ref{fig_burst_fake}) and the spectrum of persistent emission (the top panel of Figure \ref{fig_burst_fake}). The first spectrum   is given in the paragraph above. The last spectrum   is derived by stacking the spectra of the persistent emission. For the faked spectrum, we fake a spectrum of HE using the parameters derived from Figure \ref{fig_burst_spec_le_me_he},  corresponding to the burst emission which should be detected by HE if there is no deficit.

 To further estimate the variation of the deficit with the energy, two spectra are used as inputs for estimation, i.e., the detected spectrum of burst (the middle panel of Figure \ref{fig_burst_fake}), and the spectrum of persistent emission (the top panel of Figure \ref{fig_burst_fake}). The first spectrum  is given in the paragraph above. The last spectrum   is derived by stacking the spectra of the persistent emission. %For the faked spectrum, we fake a spectrum of HE using the parameters derived from Figure \ref{fig_burst_spec_le_me_he},  corresponding to the burst emission which should be detected by HE if there is no deficit.}

%{\bf  In this work, during the burst peak (the burst \#9, a blackbody model with temperature 1.86 keV and norm 0.47), the burst flux should be 0.7 cts/s  by HE in 30--100 keV (by faking a spectra of HE using XSPEC); and the burst emission at $>$ 40 keV is negligible to the persistent emission. Since the average deficit  in this energy band $\sim$7 cts/s, the burst contribution to the flux should be considered if the energy band  dips below 40 keV.}
%\textcolor{red}{JL: more discussions}
%1, Fig. 6的Y轴最好用log坐标; 2, 我也觉得不需要加这个faked spectrum。加也可以，但是不能用现在的办法，需要模拟所有暴的time-resolved的能谱，然后再加起来。

%The net deficit is the subtraction value between the faked spectrum and the detected spectrum of the burst by HE.
The fraction of the deficit is the ratio of the two spectra above, i.e., the value of the   deficit divided by the persistent emission.
As shown in Figure \ref{fig_burst_fake}, for the fraction of the deficit, there is a variability trend
%of  rising as energy.
 with energy: it increases till energy.

\section{Discussion}

This paper analyzes the broad-band X-ray lightcurves and spectra for the persistent emission and thermonuclear bursts acquired during the 2022 outburst of the  AMXP MAXI~J1816--195 observed by NICER and Insight-HXMT.

%\subsection{Comparison with  spectral results of other AMSPs}

The observations of Insight-HXMT and NICER were performed simultaneously at the peak of the outburst on 2022 June 8.
This way, we obtained a broad-band (from 1 to 100 keV) source spectrum.
We find clear evidence of a broad iron line that we interpret as produced by reflection from the inner accretion disc.
However, no significant reflection bump around 30 keV was observed.
This might be due to the high background level of ME and HE in this energy band.

The outburst's peak bolometric luminosity is roughly $<$ 30\%$L_{\rm Edd}$
  under an upper limit of the distance 6.3 kpc.
%\textcolor{red}{distance=6.3?}.
Based on the normalization of the disk emission derived from spectral fitting, the estimated inner-disk radius is $\sim$ 40 km, assuming an inclination angle of 0 (face-on scenario) and a distance of 6.3 kpc.
The inner-disk radius of AMXPs is widely used to derive the magnetic field of  the NS,  although there is a  large systematic uncertainty.
Taking the Alf\'ven radius as the magnetospheric radius \citep{Ghosh1978, Wang1997, Burderi1998}, the magnetic field is estimated as 7.1$\times10^{8}$ Gauss (\citealp{Mestel1968} or equation (4) from \citealp{DiSalvo2022}).
The co-rotation radius at which the angular velocity of Keplerian motion matches that of the NS is estimated at 25.3 km for MAXI~J1816--195.
Considering the inclination angle and distance uncertainties, the two radii above are consistent.

%depends weakly on all the parameters and is in line with the value 0.5 derived by some authors & Lamb (Ghosh 1991) More recently, has studied the torque Wang (1997) exerted on an oblique rotator by a magnetically threaded accretion disk. He concluded that / increases as the inclination angle s between the magnetic moment and the spin axis (assumed normal to the disk plane) decreases. Indeed, his estimates give /D 1 for all inclinations.

We notice that the ratio $\alpha$ between the fluxes of the persistent and bursts emission is $\sim$ 45,  which is consistent with the value derived from 15 bursts detected by  NICER  \citep{Bult2022b}.
%If we assume that all accreted matter burns during an X-ray burst, the value is close to the ratio of the relative efficiency of the two processes of hydrogen, i.e., between gravitational energy (from infinite to the NS surface) and the nuclear fusion energy \citep{Galloway}. 
%If we assume that the matter that involved  is contributes to the persistent flux. Thus the mass involved in the jet or the wind near the NS is a small value.
  Accompanying with most of the burst durations $\tau>$ 10 s and the burst occurred persistent emission $<$ 30\%$L_{\rm Edd}$, it indicates that most of these bursts occurred when  the helium ignites unstably in a hydrogen-rich environment, i.e., a mixed hydrogen/helium  burning since the hydrogen is accreted faster than it can be consumed by steady burning. 
For the last several bursts, including the bright one, the persistent emission decayed to one-third of its peak flux $<$ 10\%$L_{\rm Edd}$, and these bursts behave with a shorter duration of $\sim$ 10 s and a brighter peak flux, which indicates a change of material in the burning: a pure helium burning instead of a mixed  hydrogen/helium burning because of the hydrogen burning stably into the helium between bursts under a lower accretion rate than that of the preceding bursts.
These findings are consistent with the canonical theoretical ignition models  of  thermonuclear bursts (e.g. \citealt{Galloway}).

%The small value of $\alpha$ indicates that the almost the whole matter involved in the persistent emission has been landed on the NS surface. In other words, the mass involved in the jet (if have)  or expelled from the inner-disk near the corotation radius by the  ‘propeller’ effect is relatively small, which could be a clue to understanding the accretion process in AMXPs \citep{Lipunova2022}. }
%\textcolor{red}{still have issues here. In addition, see Frequent long bursts at very low Mdot or  Frequent long bursts at moderate Mdot in Galloway 2008}

%The small value of $\alpha$ indicates that the percentage of the outflow 

%In this case, about the same size of the magnetospheric radius and the co-rotation radius, outflow could be ignored.

%However, this radius is bigger  the magnetospheric radius.
%A 'propeller' effect should happen in this scenario, i.e., centrifugal force prevents the material entering the magnetosphere and leads to  cease of accretion.

%and thus accretion onto magnetic poles ceasesfrom entering the magnetosphere.

 For the persistent emission, we notice that the spectrum could be well fitted with a convolution thermal Comptonization model with  input seed photons from the accretion disk, and the thermal emission from the NS surface is not observed.
In particular, the scattered/covering fraction of the corona on the disk is $\sim$ 50\%, which indicates that half of the disk is covered by the corona. In this case, the disk corona model is preferred than the lamp-post geometry of the corona.

%{\bf Considering another soft X-ray generator--the accretion column of the magnetic pole--is near the NS surface and} the inner disk radius is $>$ 40 km,
%\textcolor{red}{Why?}
%这里好像和之前的版本换了意思

From the burst lightcurves of LE and ME, the burst rises tend to have a convex shape, suggesting ignition near the equator \citep{Maurer2008}.
This scenario is somewhat incompatible with the hot spot near the poles.
The inconsistency above may indicate another channel of accretion onto a magnetized NS, e.g., via Schwarzschild-Kruskal instability (the magnetic version of the Rayleigh-Taylor instability) \citep{Arons1976, Elsner1997} in the disk equatorial plane.
This model proposes that several ``tongues'' of plasma penetrate the magnetosphere and impact the NS surface at random positions.
This model has also been used to explain why bright LMXBs do not pulsate  \citep{DiSalvo2022}.%Kong: 文献？

Bursting behavior is known to be extremely variable and violent, and the bursts influence the accretion process.
%of the AMXPs are no exception to this rule.
To date, a significant soft excess is detected in  most bright bursts %with similar intensity detected by NICER,
which is caused by the interaction between the burst and the corona or/and disk.
However, in this work, this phenomenon is absent, or conservatively, not as significant as in other bursters.
Accompanying with the small fraction of the deficit in 30--100 keV, e.g., the deficit fraction of 4U~1636--536 \citep{ji2013,chen2018} and Aql~X--1 \citep{chen2013} is roughly unity,  a larger inner disk radius than other bursters is preferred,
 since the radiation pressure and the flux density of the burst imposed on the disk/corona is  proportional to 1/$R_{\rm disk}$.
Other possibilities, such as MAXJ~J1816--195 has different structures/physical-parameters of the disk/corona than other bursters can not be ruled out.  
%Kong: 这个地方没有看懂什么意思。
When more NICER data are available, more bursts from MAXI~J1816--195 will be found to be simultaneously detected by NICER and Insight-HXMT.
We expect to study those bursts in the forthcoming publication to better understand the interaction between the bursts and inferred outburst emission.

%pow(6.6710*pow(10,-8)*1.4*2*pow(10,33)/4/3.1415/3.1415/528/528,0.333)/10/1000/10
%(const double)2.53224518202494657e+01

% or different hard X-ray mechanics are

\begin{figure}[t]
\centering
\includegraphics[angle=0, scale=0.5]{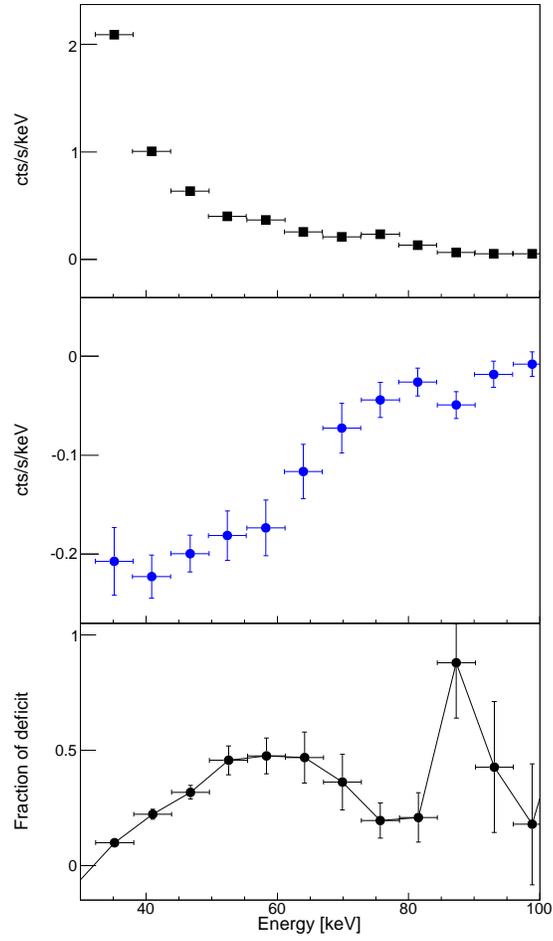}
  \caption{Top panel: the  spectrum of the persistent emission by HE. % Middle: the faked  spectrum (red) of the bursts and the detected spectrum (blue) of the bursts.
  Middle panel:  the detected spectrum  of the bursts.
  Bottom panel: deficit fraction VS energy during the bursts detected by HE.
  }
\label{fig_burst_fake}
\end{figure}

In the scenario that a burst cools the corona to a lower temperature, the fraction of the deficit should have a larger value in a higher energy band.
For instance, assuming that the corona temperature is cooled by an amplitude of 10\% during the bursts (from 10.8 keV to 9.7 keV), the count rates in 50--60 keV and 30--40 keV will be dropped by 50\% and 30\% respectively.
The above estimation is consistent with the results shown in Figure \ref{fig_burst_fake}.
We also notice that there is a clue that the  fraction of the deficit drops at $\sim$65--85 keV. This maybe relate to another hard X-ray generator--the accretion column of the magnetic pole, which is not affected by the burst.

However, this trend is not significantly possibly detected because of the low counts rate at $>85$ keV. %insufficient statistics.
It may also be the case that, at high energies, the hard X-rays are mostly from the accretion column of the magnetic pole rather than from the surrounding corona, which usually has a relatively lower temperature.
Thus, dominant hard X-rays from the magnetic pole than from the corona could, in principle, contribute to such a decreasing deficit trend at higher energies. In this case, we predict that the pulsation fraction of hard X-rays should be higher than that of soft X-rays.
%Large area X-ray missions, such as eXTP \citep{Zhang2019} will achieve more hard X-ray photons and might provide more details on the interaction between the burst and accretion process.

\bibliographystyle{plainnat}
%\bibliographystyle{natbib}

%\subsection{Conclusions and Outlook}

\acknowledgements
%We thank the reviewer for the constructive feedback and comments that greatly improved the quality of this paper.
This work made use of the data and software from the Insight-HXMT
mission, a project funded by China National Space Administration
(CNSA) and the Chinese Academy of Sciences (CAS).
This research has  made use of data and software provided by of data obtained from the High Energy Astrophysics Science
Archive Research Center (HEASARC), provided by NASA’s
Goddard Space Flight Center.
This work is supported by the National Key R\&D Program of China (2021YFA0718500) and the National Natural Science Foundation of China under grants 11733009, U1838201, U1838202, U1938101, U2038101.

% sqrt(3.78E38/4/3.14159/8.03E-8)/3.0856776E+21
%(const double)6.27238678075521960e+00

%root [9] pow(40*1000*100/(3.76E6*pow(0.13*130,-2./7)*pow(1.4,-1./7)),7./4)/1.E18*1.E26
%(const double)4.98312311813240647e+08
%root [10] pow(40*1000*100/(3.76E6*pow(0.13*200,-2./7)*pow(1.4,-1./7)),7./4)/1.E18*1.E26
%(const double)6.18080353382901788e+08
%pow(40*1000*100/(3.76E6*pow(0.289*200,-2./7)*pow(1.4,-1./7)),7./4)/1.E18*1.E26
%(const double)9.21556961868013859e+08

%pow(3.96E6/(3.7E6*pow(0.289*1.3*100,-2./7.)*pow(1.4,-1./7.)),7./4.)/1.E18*1.E26
%(const double)7.50872440717788577e+08

%root [1] 0.63*sqrt(5010)
%(const double)4.45922526903496959e+01
%root [2] pow(44.6*1000*100/(3.76E6*pow(0.13*130,-2./7)*pow(1.4,-1./7)),7./4)/1.E18*1.E26
%(const double)6.02882441224476576e+08
%root [3] 0.63/2/sqrt(5010)*2820
%(const double)1.25499154278229703e+01
%root [4] 0.63/2/sqrt(5010)*1540
%(const double)6.85349991448488449e+00

% 1.11E-8*4.*3.14159*pow(6.3*3.08E21,2)/1.3/1.4/1.E38
%(const double)2.88564977044201865e-01 Ledd
%1.11E-8*4.*3.14159*pow(6.3*3.08E21,2)/3.8E38
%(const double)1.38207436373801951e-01 Ledd

\begin{table}\tiny
\centering
\caption{The results of the spectral fitting of the joint LE, ME, HE and NICER spectrum in the 1--100 keV range   with cons*tbabs*(thcomp*diskbb+gauss)}
\label{tb_persist_fit}
%\vskip -0.4cm
\begin{tabular}{ccccccccccc}
\\\hline
$N_{\rm H}$  & $\tau$ & $kT_{\rm e}$  & $f_{\rm sc}$ & $kT_{\rm in}$
 & $N_{\rm diskbb}$&  $E_{\rm Fe}$ & $\sigma_{\rm Fe}$ &$A_{\rm Fe}$  & $\chi_\nu^2$  &$F_{\rm bol}$\\
$10^{22}~{\rm cm}^{-2}$& & keV  &keV & & $10^2$ & keV & keV & $10^{-3}$~cts/${\rm cm}^{2}/{\rm s} $ & & $10^{-8}~{\rm erg/cm}^{2}/{\rm s}$\\\hline
%$2.81_{-0.21}^{+0.22}$ & $5.19^{+0.23}_{-0.24}$ & $11.4^{+0.7}_{-0.6}$ & $0.46_{-0.05}^{+0.05}$  & $0.47_{-0.02}^{+0.03}$ & $50.1_{-15.4}^{+28.2}$ & 6.4 (fxd)& $0.90_{-0.18}^{+0.17}$ & $5.7_{-1.5}^{+1.6}$ & 215/206  &$1.19_{-0.01}^{+0.01}$\\ %0.4-100 keV result
$2.37_{-0.03}^{+0.02}$ & $5.41^{+0.25}_{-0.07}$ & $10.8^{+0.6}_{-0.2}$ & $0.52_{-0.01}^{+0.02}$  & $0.48_{-0.01}^{+0.01}$ & $39.5_{-5.1}^{+5.1}$ & 6.4 (fxd)& $0.86_{-0.14}^{+0.15}$ & $5.4_{-1.1}^{+0.9}$ & 217/192  &$1.11_{-0.01}^{+0.01}$\\\hline %1-100 keV result
\end{tabular}
%\begin{list}{}{}
%\item[Note:]{ }
%\end{list}
\end{table}

\clearpage

%\begin{center}
\centering
%\caption{The bursts detected by Insight/HXMT in 2022 outburst of MAXI~J1816--195}
\begin{longtable}{cccccc}%\tiny
\caption{The bursts detected by Insight/HXMT in 2022 outburst of MAXI~J1816--195}
\label{tb_burst_fit}\\
\hline
 No & OBSID &Time   & $F_{\rm p}$ & $E_{\rm b}$ & $\tau$ \\
  &  &MJD &  $10^{-8}~{\rm erg/cm}^{2}/{\rm s}$ &  $10^{-8}~{\rm erg/cm}^{2}$ & s \\\hline
1 & P040427500101-20220608-01-01 & 59738.23551 & $3.17_{-0.41}^{+0.47}$ & $61.9_{-2.5}^{+2.6}$ & $19.5_{-2.6}^{+3.0}$ \\ \hline
2 & P040427500101-20220608-01-01 & 59738.31022 & $3.02_{-0.34}^{+0.41}$ & $59.8_{-2.2}^{+2.3}$ & $19.8_{-2.4}^{+2.8}$ \\ \hline
3 & P040427500102-20220608-01-01 & 59738.38172 & $3.02_{-0.39}^{+0.47}$ & $73.6_{-2.9}^{+3.1}$ & $24.3_{-3.3}^{+3.9}$ \\ \hline
4 & P040427500102-20220608-01-01 & 59738.45275 & $3.45_{-0.37}^{+0.43}$ & $60.5_{-2.4}^{+2.6}$ & $17.5_{-2.0}^{+2.3}$ \\ \hline
5 & P040427500103-20220608-01-01 & 59738.52387 & $3.52_{-0.44}^{+0.52}$ & $70.8_{-2.6}^{+2.8}$ & $20.1_{-2.6}^{+3.1}$ \\ \hline
6 & P040427500103-20220608-01-01 & 59738.59306 & $3.43_{-0.45}^{+0.54}$ & $71.2_{-2.9}^{+3.1}$ & $20.7_{-2.9}^{+3.4}$ \\ \hline
7 & P040427500104-20220608-01-01 & 59738.66072 & $4.14_{-0.50}^{+0.59}$ & $70.0_{-3.1}^{+3.3}$ & $16.9_{-2.2}^{+2.5}$ \\ \hline
8 & P040427500104-20220608-01-01 & 59738.72614 & $3.60_{-0.16}^{+0.16}$ & $88.1_{-1.0}^{+1.0}$ & $24.5_{-1.1}^{+1.1}$ \\ \hline
9${\mathrm{*}}$ & P040427500105-20220608-01-01 & 59738.78847 & $3.54_{-0.17}^{+0.17}$ & $96.6_{-1.1}^{+1.2}$ & $27.3_{-1.3}^{+1.3}$ \\ \hline
10 & P040427500105-20220608-01-01 & 59738.85032 & $4.27_{-0.53}^{+0.63}$ & $75.1_{-3.1}^{+3.3}$ & $17.6_{-2.3}^{+2.7}$ \\ \hline
11 & P040427500107-20220608-01-01 & 59739.02920 & $4.19_{-0.58}^{+0.71}$ & $70.5_{-2.4}^{+2.6}$ & $16.8_{-2.4}^{+2.9}$ \\ \hline
12 & P040427500201-20220609-01-01 & 59739.19766 & $3.40_{-0.41}^{+0.47}$ & $76.7_{-3.0}^{+3.2}$ & $22.6_{-2.9}^{+3.3}$ \\ \hline
13 & P040427500202-20220609-01-01 & 59739.30639 & $3.00_{-0.32}^{+0.36}$ & $77.9_{-3.0}^{+3.2}$ & $25.9_{-2.9}^{+3.3}$ \\ \hline
14 & P040427500202-20220609-01-01 & 59739.36110 & $3.44_{-0.42}^{+0.49}$ & $73.1_{-2.7}^{+2.9}$ & $21.2_{-2.7}^{+3.1}$ \\ \hline
15 & P040427500204-20220609-01-01 & 59739.58436 & $3.60_{-0.55}^{+0.68}$ & $80.3_{-3.1}^{+3.3}$ & $22.3_{-3.5}^{+4.3}$ \\ \hline
16 & P040427500204-20220609-01-01 & 59739.64061 & $3.38_{-0.40}^{+0.48}$ & $81.6_{-3.2}^{+3.4}$ & $24.1_{-3.0}^{+3.6}$ \\ \hline
17 & P040427500205-20220609-01-01 & 59739.69725 & $2.90_{-0.31}^{+0.35}$ & $72.2_{-2.7}^{+2.9}$ & $24.9_{-2.8}^{+3.2}$ \\ \hline
18 & P040427500205-20220609-01-01 & 59739.75401 & $3.82_{-0.49}^{+0.58}$ & $74.6_{-2.9}^{+3.0}$ & $19.5_{-2.6}^{+3.1}$ \\ \hline
19 & P040427500206-20220609-01-01 & 59739.92397 & $3.43_{-0.39}^{+0.45}$ & $77.8_{-3.0}^{+3.1}$ & $22.7_{-2.7}^{+3.1}$ \\ \hline
20 & P040427500207-20220609-01-01 & 59739.97869 & $3.58_{-0.17}^{+0.17}$ & $100.9_{-1.1}^{+1.1}$ & $28.2_{-1.3}^{+1.3}$ \\ \hline
21 & P040427500302-20220610-01-01 & 59740.23140 & $2.74_{-0.29}^{+0.33}$ & $73.0_{-2.8}^{+2.9}$ & $26.7_{-3.0}^{+3.4}$ \\ \hline
22 & P040427500303-20220610-01-01 & 59740.43116 & $3.42_{-0.50}^{+0.64}$ & $74.2_{-3.0}^{+3.2}$ & $21.7_{-3.3}^{+4.2}$ \\ \hline
23 & P040427500304-20220610-01-01 & 59740.48185 & $3.19_{-0.41}^{+0.48}$ & $70.2_{-2.7}^{+2.9}$ & $22.0_{-2.9}^{+3.5}$ \\ \hline
24 & P040427500304-20220610-01-01 & 59740.58205 & $3.28_{-0.16}^{+0.16}$ & $82.9_{-1.1}^{+1.1}$ & $25.3_{-1.2}^{+1.2}$ \\ \hline
25 & P040427500305-20220610-01-01 & 59740.62985 & $2.87_{-0.50}^{+0.66}$ & $61.5_{-2.8}^{+3.0}$ & $21.5_{-3.9}^{+5.0}$ \\ \hline
26 & P040427500305-20220610-01-01 & 59740.68092 & $3.20_{-0.38}^{+0.45}$ & $75.0_{-2.9}^{+3.1}$ & $23.4_{-2.9}^{+3.4}$ \\ \hline
27 & P040427500306-20220610-01-01 & 59740.78507 & $2.60_{-0.30}^{+0.36}$ & $68.0_{-2.6}^{+2.8}$ & $26.1_{-3.2}^{+3.7}$ \\ \hline
28 & P040427500306-20220610-01-01 & 59740.83742 & $3.42_{-0.17}^{+0.17}$ & $94.5_{-1.1}^{+1.1}$ & $27.6_{-1.4}^{+1.4}$ \\ \hline
29 & P040427500307-20220610-01-01 & 59740.88945 & $3.58_{-0.55}^{+0.67}$ & $75.1_{-3.2}^{+3.4}$ & $21.0_{-3.3}^{+4.0}$ \\ \hline
30 & P040427500401-20220611-01-01 & 59741.04474 & $3.54_{-0.16}^{+0.16}$ & $92.8_{-1.1}^{+1.1}$ & $26.2_{-1.2}^{+1.2}$ \\ \hline
31 & P040427500401-20220611-01-01 & 59741.14988 & $3.35_{-0.40}^{+0.47}$ & $77.9_{-2.9}^{+3.1}$ & $23.2_{-2.9}^{+3.4}$ \\ \hline
32 & P040427500403-20220611-01-01 & 59741.36075 & $2.68_{-0.26}^{+0.29}$ & $62.6_{-2.3}^{+2.5}$ & $23.3_{-2.4}^{+2.6}$ \\ \hline
33 & P040427500403-20220611-01-01 & 59741.41406 & $3.23_{-0.35}^{+0.39}$ & $74.8_{-2.9}^{+3.1}$ & $23.1_{-2.6}^{+3.0}$ \\ \hline
34 & P040427500404-20220611-01-01 & 59741.57413 & $3.41_{-0.16}^{+0.16}$ & $90.1_{-1.1}^{+1.1}$ & $26.4_{-1.3}^{+1.3}$ \\ \hline
35 & P040427500405-20220611-01-01 & 59741.62791 & $2.78_{-0.33}^{+0.39}$ & $69.6_{-2.7}^{+2.8}$ & $25.0_{-3.1}^{+3.6}$ \\ \hline
36 & P040427500405-20220611-01-01 & 59741.68257 & $3.13_{-0.47}^{+0.62}$ & $75.4_{-3.0}^{+3.1}$ & $24.1_{-3.7}^{+4.9}$ \\ \hline
37 & P040427500407-20220611-01-01 & 59741.90167 & $3.61_{-0.16}^{+0.16}$ & $84.1_{-0.9}^{+0.9}$ & $23.3_{-1.1}^{+1.1}$ \\ \hline
38 & P040427500501-20220612-01-01 & 59742.23829 & $4.22_{-0.59}^{+0.72}$ & $74.8_{-3.9}^{+4.2}$ & $17.7_{-2.6}^{+3.2}$ \\ \hline
39 & P040427500501-20220612-01-01 & 59742.35230 & $3.55_{-0.34}^{+0.38}$ & $60.1_{-2.2}^{+2.3}$ & $16.9_{-1.7}^{+1.9}$ \\ \hline
40 & P040427500503-20220612-01-01 & 59742.63528 & $3.73_{-0.16}^{+0.16}$ & $98.4_{-1.0}^{+1.1}$ & $26.4_{-1.2}^{+1.2}$ \\ \hline
41 & P040427500504-20220612-01-01 & 59742.69325 & $3.49_{-0.16}^{+0.16}$ & $84.3_{-0.9}^{+0.9}$ & $24.1_{-1.1}^{+1.1}$ \\ \hline
42 & P040427500504-20220612-01-01 & 59742.75051 & $3.55_{-0.53}^{+0.66}$ & $70.1_{-2.8}^{+2.9}$ & $19.7_{-3.1}^{+3.8}$ \\ \hline
43 & P040427500505-20220612-01-01 & 59742.80863 & $4.22_{-0.47}^{+0.54}$ & $72.3_{-2.4}^{+2.5}$ & $17.1_{-2.0}^{+2.3}$ \\ \hline
44 & P040427500505-20220612-01-01 & 59742.86578 & $3.65_{-0.46}^{+0.54}$ & $75.3_{-2.8}^{+2.9}$ & $20.6_{-2.7}^{+3.2}$ \\ \hline
45 & P040427500601-20220613-01-01 & 59743.09808 & $3.42_{-0.16}^{+0.16}$ & $95.3_{-1.0}^{+1.0}$ & $27.9_{-1.3}^{+1.3}$ \\ \hline
46 & P040427500602-20220613-01-01 & 59743.27835 & $3.66_{-0.43}^{+0.50}$ & $75.8_{-2.8}^{+2.9}$ & $20.7_{-2.5}^{+2.9}$ \\ \hline
47 & P040427500602-20220613-01-01 & 59743.33933 & $3.53_{-0.48}^{+0.58}$ & $77.7_{-3.1}^{+3.3}$ & $22.0_{-3.1}^{+3.7}$ \\ \hline
48 & P040427500603-20220613-01-01 & 59743.40185 & $3.26_{-0.36}^{+0.42}$ & $75.5_{-2.8}^{+2.9}$ & $23.2_{-2.7}^{+3.1}$ \\ \hline
49 & P040427500603-20220613-01-01 & 59743.46334 & $3.26_{-0.40}^{+0.48}$ & $63.4_{-2.1}^{+2.2}$ & $19.5_{-2.5}^{+3.0}$ \\ \hline
50 & P040427500607-20220613-01-01 & 59743.96283 & $3.45_{-0.41}^{+0.48}$ & $77.4_{-2.7}^{+2.9}$ & $22.4_{-2.8}^{+3.2}$ \\ \hline
51 & P040427500607-20220613-01-01 & 59744.02764 & $3.41_{-0.42}^{+0.50}$ & $79.1_{-2.8}^{+3.0}$ & $23.2_{-3.0}^{+3.5}$ \\ \hline
52 & P040427500608-20220614-02-01 & 59744.09221 & $3.87_{-0.16}^{+0.16}$ & $101.6_{-1.0}^{+1.0}$ & $26.2_{-1.1}^{+1.1}$ \\ \hline
53 & P040427500608-20220614-02-01 & 59744.15880 & $3.71_{-0.16}^{+0.16}$ & $97.5_{-1.0}^{+1.0}$ & $26.3_{-1.2}^{+1.2}$ \\ \hline
54 & P040427500609-20220614-02-01 & 59744.22394 & $3.77_{-0.17}^{+0.17}$ & $89.5_{-1.0}^{+1.0}$ & $23.7_{-1.1}^{+1.1}$ \\ \hline
55 & P040427500609-20220614-02-01 & 59744.28849 & $3.69_{-0.50}^{+0.61}$ & $112.6_{-1.8}^{+1.8}$ & $30.6_{-4.1}^{+5.1}$ \\ \hline
56 & P040427500611-20220614-02-01 & 59744.48935 & $4.00_{-0.16}^{+0.16}$ & $111.1_{-1.0}^{+1.0}$ & $27.8_{-1.1}^{+1.1}$ \\ \hline
57 & P040427500611-20220614-02-01 & 59744.55683 & $3.34_{-0.41}^{+0.48}$ & $76.3_{-2.8}^{+3.0}$ & $22.9_{-2.9}^{+3.4}$ \\ \hline
58 & P040427500612-20220614-02-01 & 59744.62540 & $3.30_{-0.44}^{+0.53}$ & $76.9_{-3.2}^{+3.4}$ & $23.3_{-3.2}^{+3.9}$ \\ \hline
59 & P040427500701-20220615-01-01 & 59745.52358 & $3.57_{-0.42}^{+0.48}$ & $79.6_{-2.9}^{+3.1}$ & $22.3_{-2.7}^{+3.1}$ \\ \hline
60 & P040427500701-20220615-01-01 & 59745.59710 & $3.67_{-0.44}^{+0.52}$ & $81.7_{-3.4}^{+3.6}$ & $22.2_{-2.8}^{+3.3}$ \\ \hline
61 & P040427500702-20220615-01-01 & 59745.67103 & $4.19_{-0.53}^{+0.64}$ & $75.2_{-3.0}^{+3.1}$ & $17.9_{-2.4}^{+2.8}$ \\ \hline
62 & P040427500702-20220615-01-01 & 59745.74569 & $3.96_{-0.16}^{+0.16}$ & $97.0_{-1.0}^{+1.0}$ & $24.5_{-1.0}^{+1.0}$ \\ \hline
63 & P040427500801-20220616-01-01 & 59746.66017 & $3.25_{-0.45}^{+0.53}$ & $65.3_{-2.7}^{+2.9}$ & $20.1_{-2.9}^{+3.4}$ \\ \hline
64 & P040427500801-20220616-01-01 & 59746.74048 & $3.85_{-0.16}^{+0.16}$ & $98.3_{-1.0}^{+1.0}$ & $25.6_{-1.1}^{+1.1}$ \\ \hline
65 & P040427500901-20220617-01-01 & 59747.73360 & $3.58_{-0.39}^{+0.46}$ & $79.7_{-3.0}^{+3.1}$ & $22.3_{-2.6}^{+3.0}$ \\ \hline
66 & P040427500903-20220617-01-01 & 59747.99203 & $3.94_{-0.16}^{+0.16}$ & $98.9_{-1.0}^{+1.0}$ & $25.1_{-1.1}^{+1.1}$ \\ \hline
67 & P040427501101-20220619-01-01 & 59749.36886 & $4.24_{-0.46}^{+0.52}$ & $66.0_{-2.4}^{+2.5}$ & $15.6_{-1.8}^{+2.0}$ \\ \hline
68 & P040427501201-20220620-01-01 & 59750.41864 & $3.92_{-0.43}^{+0.50}$ & $63.3_{-2.4}^{+2.6}$ & $16.2_{-1.9}^{+2.2}$ \\ \hline
69 & P040427501301-20220621-01-01 & 59751.30502 & $4.08_{-0.17}^{+0.17}$ & $78.1_{-0.9}^{+0.9}$ & $19.1_{-0.8}^{+0.8}$ \\ \hline
70 & P040427501301-20220621-01-01 & 59751.41726 & $3.71_{-0.40}^{+0.46}$ & $63.5_{-2.6}^{+2.8}$ & $17.1_{-2.0}^{+2.3}$ \\ \hline
71 & P040427501901-20220626-01-01 & 59756.85801 & $6.60_{-0.22}^{+0.22}$ & $87.7_{-0.9}^{+0.9}$ & $13.3_{-0.5}^{+0.5}$ \\ \hline
72 & P040427502001-20220627-01-01 & 59757.65185 & $4.84_{-0.38}^{+0.42}$ & $52.8_{-1.8}^{+1.9}$ & $10.9_{-0.9}^{+1.0}$ \\ \hline
73 & P040427502101-20220628-01-01 & 59758.58367 & $8.03_{-0.39}^{+0.40}$ & $82.6_{-1.5}^{+1.5}$ & $10.3_{-0.5}^{+0.5}$ \\ \hline
%\end{tabular}
\end{longtable}
%\end{center}
\begin{list}{}{}
\item[${\mathrm{*}}$]{The burst is detected simultaneously by NICER\&Insight-HXMT}
\end{list}
%\end{table}
\clearpage

%\begin{figure}[t]
%\centering
%\includegraphics[angle=-90, scale=0.5]{10.eps}
%  \caption{The  spectral fits results by NICER (black), Insight-HXMT/LE (red) and Insight-HXMT/ME when the burst reach the maximum emission area by an absorbed black-body  model.  }
%\label{fig_burst_spec}
%\end{figure}

%\begin{figure}[t]
%\centering
%\includegraphics[angle=-90, scale=0.5]{xspec_le_me_he_fixed_2.eps}
%  \caption{The stacked burst spectrum fitted with model cons*wabs*(bbodryad+bbodyrad). The energy ranges of the spectrum of LE (black), ME (red) and HE (green) are 1--10 keV, 8–30 keV and 30–100 keV, respectively. Please note that only the LE data and the ME data are used in the spectral fitting, the HE data are not involved because of their negative values; i.e., the HE data are just for display purposes--indicating the hard X-ray shortage. %  since the negative value cannot been shown in the main panel with a logarithmic coordinate axis  }
%\label{fig_burst_spec_le_me_he}
%\end{figure}

%\begin{figure}[t]
%\centering
%\includegraphics[angle=0, scale=0.5]{F_b_F_p.eps}
%  \caption{Deficit fraction VS energy during bursts detected by HE.  }
%\label{fig_burst_fake}
%\end{figure}
\end{document}